\documentclass[a4paper,fleqn]{cas-dc}

\usepackage[numbers, sort&compress]{natbib}
\usepackage{lipsum}
\usepackage[flushleft]{threeparttable}
\usepackage{xurl}

\def\tsc#1{\csdef{#1}{\textsc{\lowercase{#1}}\xspace}}
\tsc{WGM}
\tsc{QE}

\begin{document}
\let\WriteBookmarks\relax
\def\floatpagepagefraction{1}
\def\textpagefraction{.001}

\shorttitle{The nature of silicon PN junction impedance at high frequency}    

\shortauthors{D.A. van Nijen et al.}  

\title [mode = title]{The nature of silicon PN junction impedance at high frequency} 



%

\author[1]{David A. van Nijen}[orcid=0000-0002-2235-0285]

\cormark[1]


\ead{d.a.vannijen@tudelft.nl}


\credit{Conceptualization; Data curation; Formal analysis; Investigation; Methodology; Project administration; Software; Validation; Visualization; Writing - original draft; Writing - Review \& Editing}

\affiliation[1]{organization={Delft University of Technology},
            addressline={Mekelweg 4}, 
            city={Delft},
            postcode={2628 CD}, 
            country={The Netherlands}}
\author[1]{Paul Procel}  \credit{Methodology}
\author[1]{René A.C.M.M. van Swaaij}  \credit{Writing - Review \& Editing}
\author[1]{Miro Zeman}  \credit{Funding Acquisition}
\author[1]{Olindo Isabella} \credit{Funding Acquisition; Supervision; Writing - Review \& Editing}
\author[1]{Patrizio Manganiello}[]  \credit{Conceptualization; Formal analysis; Methodology; Project administration; Software; Supervision; Writing - Review \& Editing}


\cortext[1]{Corresponding author}



\begin{abstract}
A thorough understanding of the small-signal response of solar cells can reveal intrinsic device characteristics and pave the way for innovations. 
This study investigates the impedance of crystalline silicon PN junction devices using TCAD simulations, focusing on the impact of frequency, bias voltage, and the presence of a low-high (LH) junction. 
It is shown that the PN junction exhibits a fixed $RC$-loop behavior at low frequencies, but undergoes relaxation in both resistance $R_{j}$ and capacitance $C_{j}$ as frequency increases. Moreover, it is revealed that the addition of a LH junction impacts the impedance by altering $R_{j}$, $C_{j}$, and the series resistance $R_{s}$. Contrary to conventional modeling approaches, which often include an additional $RC$-loop to represent the LH junction, this study suggests that such a representation does not represent the underlying physics, particularly the frequency-dependent behavior of $R_{j}$ and $C_{j}$.
\end{abstract}




\maketitle



\section{Introduction}
In photovoltaic (PV) systems, solar cells are commonly perceived as devices that produce direct current (DC) upon exposure to sunlight. Much of the research and development in solar energy focuses on enhancing the efficiency of solar cells in converting light into electrical power. However, the potential use of solar cells for high-frequency alternating current (AC) applications makes it valuable to explore and refine their impedance.
For instance, studying the solar-cell impedance can rapidly gather large amounts of data for efficient and accurate measurements of a variety of solar cell parameters \cite{Mora2009,Garland2011a,MoraSero2008}. 
Additionally, the solar cell impedance influences the performance of power electronics that perform maximum power point tracking (MPPT) \cite{Panigrahi2016}. Extending this concept, it has been proposed to exploit the self-impedance of solar cells, which could lead to the development of converters that require fewer passive components at the input \cite{vannijen2022,Huang2016}.
Moreover, the impedance of solar cells impacts their suitability for applications beyond energy generation, such as Visible Light Communication (VLC) systems. In such systems, where solar cells can be employed as receivers \cite{Ziar2021,kim2014simultaneous,RSarwar2017}, the impedance directly influences the system's bandwidth \cite{YZhou2023}.
These examples highlight the potential of solar-cell impedance analysis to reveal intrinsic device characteristics and pave the way for innovations. 

In the context of small-signal analysis, it is well known that the PN junction impedance cannot be fully represented by any linear, lumped, finite, passive, or bilateral network \cite{Neamen2012}. Nevertheless, when the frequency is sufficiently low, the PN junction impedance can be approximated by a parallel resistor-capacitor circuit ($RC$-loop), a common simplification found in literature. At these lower frequencies, the minority carrier distribution can effectively "follow" the frequency of the AC signal. Yet, as the frequency increases beyond a certain threshold, violating this condition, the $RC$-loop approximation loses its validity. 
However, it is important to mention that solar cells consist of more than just a PN junction. For example, direct metal contact to the lightly doped substrate is not appropriate, as it would lead to the formation of a Schottky barrier and/or Fermi level pinning \cite{Allen2019}. Hence, practical devices often possess a so-called low-high (LH) junction, which reduces the contact resistance by facilitating tunneling and reduces surface recombination. However, this poses a challenge when the impedance of such a device no longer behaves like a single $RC$-loop beyond a certain frequency. Is this because the $RC$-loop approximation of the PN junction is no longer valid, or is it due to separate impedance effects in the LH junction? 
In solar-cell literature, various high-frequency effects are attributed to this LH junction. For instance, an additional $RC$-loop is frequently incorporated into the impedance model to account for the LH junction impedance \cite{Garland2011a,Crain2012,vannijen2023,Olayiwola2020,Panigrahi2016,Yadav2017}. Moreover, sometimes a constant phase element (CPE) or a third $RC$-loop is even introduced to solar-cell impedance models to address various non-ideal effects in devices with passivating contacts \cite{Shehata2023,Panigrahi2023}. While these adjustments often improve fitting accuracy, the underlying physics may not be accurately captured.

In this study, the impedance of crystalline silicon (c-Si) PN junction devices is investigated through Technology Computer Aided Design (TCAD) simulations. The TCAD simulation methodology enables the isolation of the PN junction impedance, since ideal contacts can be positioned on a bulk material without the inclusion of a LH junction. Moreover, the almost noise-free simulation data facilitate a thorough examination of subtle shortcomings of equivalent model fits.
While much of the theory in this paper also applies to devices with passivating contacts, this study exclusively focuses on homojunction architectures. It is crucial for the field to understand the limitations of modeling PN homojunction impedance across different frequency ranges with a resistor-capacitor circuit. Without this knowledge, non-idealities may be incorrectly attributed to other parts of the device. For instance, one outcome of this paper is that certain impedance behaviour that is often attributed to capacitive effects in the LH junction, is in fact due to non-ideal high-frequency effects in the PN junction.
First, the paper demonstrates the frequency range within which the PN junction can be accurately represented by an $RC$-loop, and it explores the consequences of exceeding this frequency threshold. Subsequently, it is demonstrated how the silicon bulk properties influence these results. Finally, it is discussed how the impedance changes when a LH junction is added to the device structure.

\section{Theory}
In section \ref{sec:PNadmittance}, the theory of PN junction admittance related to minority carrier diffusion is summarized. Subsequently, section \ref{sec:impedancecircuit} outlines a more comprehensive small-signal impedance model of a PN junction, and provides equations describing the circuit element values.

\subsection{PN junction small-signal admittance} \label{sec:PNadmittance}
When a PN junction diode is forward biased, diffusion mechanisms play an important role in the admittance. In a PN junction, the electron diffusion current is given by \cite{Neamen2012}:
\begin{equation}
    I_{n0} = \frac{A q D_{n} n_{p0}}{L_{n}} \text{exp} \left ( \frac{q V_{F}}{kT} \right )
\end{equation}
where $A$ is the cross-sectional area of the PN junction, $q$ is the elementary charge constant, $D_{n}$ is the diffusion constant of electrons, $n_{p0}$ is the minority carrier electron concentration on the p side of the junction, $L_{n}$ is the electron diffusion length, and $V_{F}$ is the applied forward-bias voltage.  Finally, $k$ and $T$ are the Boltzmann constant and temperature, respectively. Similarly, the hole diffusion current in a PN junction is given by \cite{Neamen2012}:
\begin{equation}
    I_{p0} = \frac{A q D_{p} p_{n0}}{L_{p}} \text{exp} \left ( \frac{q V_{F}}{kT} \right )
\end{equation}
where $D_{p}$ is the diffusion constant of holes, $p_{n0}$ is the minority carrier hole concentration on the n side of the junction, and $L_{p}$ is the hole diffusion length. Using these diffusion current expressions, the PN junction admittance can be defined as \cite{Neamen2012}:
\begin{equation}
    Y = \frac{1}{V_{t}} \left [I_{p0} \sqrt{1+i \omega \tau_{p0}} + I_{n0} \sqrt{1+i \omega \tau_{n0}} \right ]
    \label{eq:admittance_general}
\end{equation}
where the thermal voltage $V_{t}$ is $kT/q$, $i$ denotes the imaginary unit, $\tau_{p0}$ and $\tau_{n0}$ are the minority charge carrier lifetimes, and $\omega$ is the radian frequency of the small signal. It is worth noting that the real and imaginary terms in Equation \ref{eq:admittance_general} can be separated using the following equation:

\begin{equation}
    \sqrt{1+i \omega \tau_{0}} = \pm \left [ \sqrt{\frac{\sqrt{1+\omega^{2} \tau_{0}^{2}}+1}{2}} + i\sqrt{\frac{\sqrt{1+\omega^{2} \tau_{0}^{2}}-1}{2}} \right ]
    \label{eq:admittance_math}
\end{equation}
No linear, lumped, finite, passive, bilateral network exists that can be synthesized to give the admittance function of Equation \ref{eq:admittance_general} \cite{Neamen2012}. However, assuming that the frequency is low enough that the minority carrier distribution in the quasi-neutral regions of the junction can follow the AC voltage ($\omega \tau_{0}\ll1$), the approximation $\sqrt{1+i \omega \tau_{0}} \approx  1 + i \omega \tau_{0}/2$ can be made. Substituting this into Equation \ref{eq:admittance_general} yields the following expression for the low-frequency admittance:

\begin{equation}
    Y = \frac{1}{V_{t}} \left [ I_{p0} \left ( 1 + \frac{i \omega \tau_{p0}}{2}  \right ) + I_{n0} \left ( 1 + \frac{i \omega \tau_{n0}}{2}  \right ) \right ]
    \label{eq:admittance_lowf}
\end{equation}

As opposed to the admittance in Equation \ref{eq:admittance_general}, the low-frequency admittance in Equation \ref{eq:admittance_lowf} can be described by a lumped passive circuit. Indeed, Equation \ref{eq:admittance_lowf} can be written in the form:

\begin{equation}
    Y = g_{\text{dif}} + i \omega C_{\text{dif}}
\end{equation}
with the diffusion conductance $g_{\text{dif}}$ and diffusion capacitance $C_{\text{dif}}$ both being frequency-independent. However, Equations \ref{eq:admittance_general} and \ref{eq:admittance_math} dictate that as a certain frequency threshold is exceeded, both the real and imaginary part of the admittance are expected to increase with frequency.

\subsection{Small-signal model} \label{sec:impedancecircuit}
\begin{figure}[ht]
      \centering
      \includegraphics[width=0.45\textwidth]{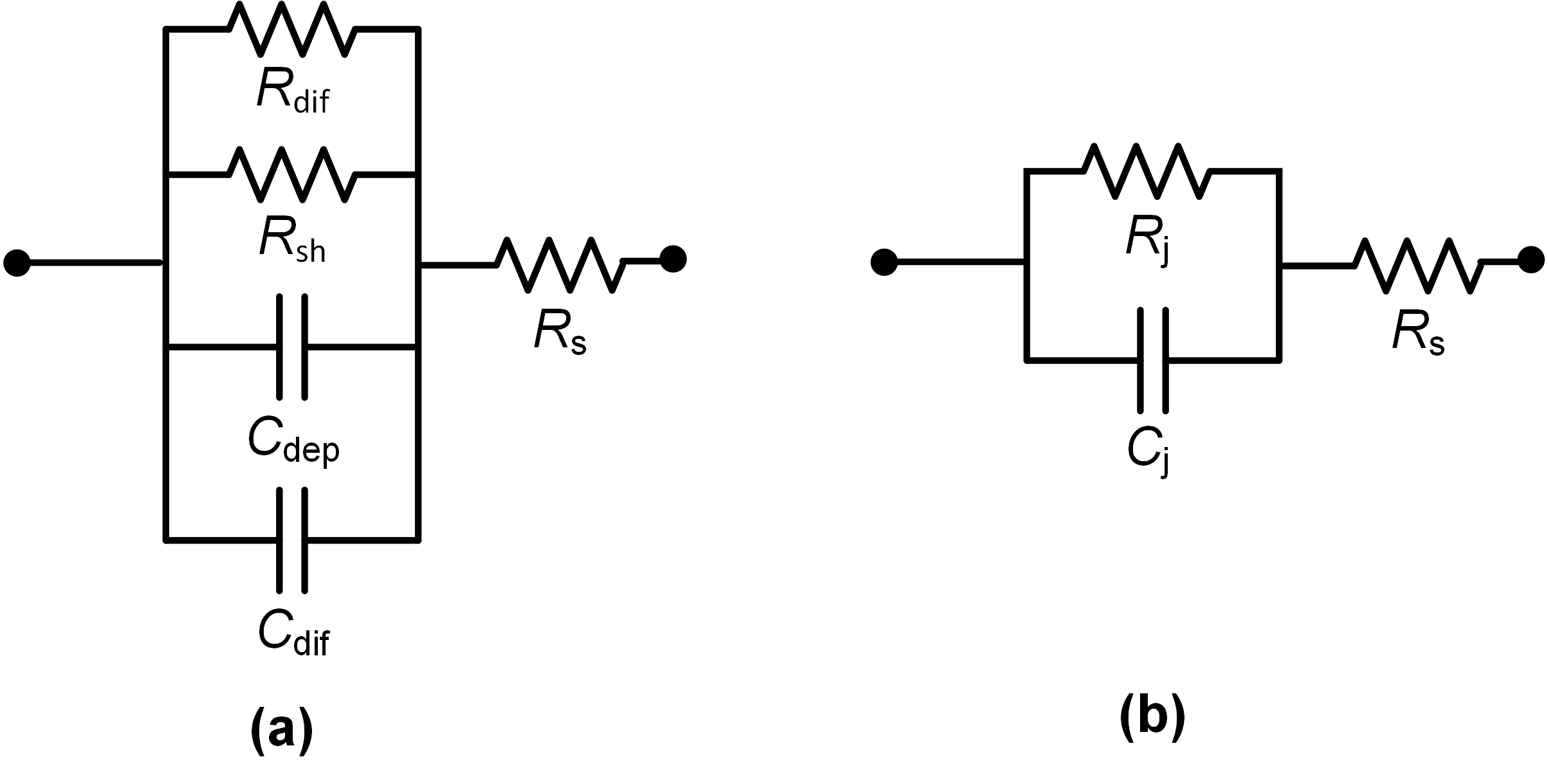}
      \caption{PN junction equivalent circuits: (a) full circuit, (b) simplified circuit used for the analysis in this work.}
      \label{fig:PN_eqcircuit}
\end{figure}

In the previous subsection it was shown that the small-signal equivalent circuit of a PN junction should include a diffusion resistance $R_{\text{dif}}$ and a diffusion capacitance $C_{\text{dif}}$ to account for the dynamics involved in the diffusion of minority carriers. Additionally, to create a more comprehensive small-signal equivalent circuit of a PN junction, other elements are incorporated. These are a shunt resistance $R_{sh}$ (whose ideal value is very high) and a series resistance $R_{s}$. Furthermore, a depletion layer capacitance $C_{\text{dep}}$ is introduced to account for the capacitance related to the ionized atoms in the depletion region. According to literature, $C_{\text{dep}}$ is not expected to relax with frequency \cite{Lucia1993}. The complete small-signal equivalent circuit incorporating all these components is presented in Figure \ref{fig:PN_eqcircuit}(a). For further analysis, this model can be simplified to the one in Figure \ref{fig:PN_eqcircuit}(b), in which $R_{j}$ is the parallel of $R_{\text{dif}}$ and $R_{sh}$, and $C_{j}$ is the parallel of $C_{\text{dep}}$ and $C_{\text{dif}}$.

The depletion layer capacitance ($C_{\text{dep}}$) typically dominates $C_{j}$ at reverse bias and at low forward bias voltage. For a PN junction consisting of homogeneously doped regions, $C_{\text{dep}}$ is given by \cite{Neamen2012}:
\begin{equation}
    C_{\text{dep}} = A \sqrt{  \frac{q \epsilon_{s} N_{a} N_{d}}{ 2 \left ( V_{bi} + V_{R} \right ) \left( N_{a} + N_{d} \right ) }}
    \label{eq:junccap}
\end{equation}
where $\epsilon_{s}$ is the permittivity of the semiconductor, $N_{a}$ and $N_{d}$ are the acceptor and donor concentrations, respectively. $V_{bi}$ is the built-in potential, and $V_{R}$ is the applied reverse-bias voltage. 
The low-frequency diffusion capacitance ($C_{\text{dif}}$) typically dominates $C_{j}$ at high forward bias voltage. For a PN junction consisting of homogeneously doped regions, $C_{\text{dif}}$ is given by \cite{Neamen2012}:
\begin{equation}
    C_{\text{dif}} =  \frac{q^{2} n_{i}^{2} A }{2kT}  \left ( \frac{\sqrt{D_{p} \tau_{p0}}}{N_{d}}  + \frac{\sqrt{D_{n} \tau_{n0}}}{N_{a}}  \right ) \text{exp} \left (\frac{q V_{F}} {kT}  \right )
    \label{eq:diffcap}
\end{equation}
where $n_{i}$ is the intrinsic carrier concentration and $V_{F}$ is the applied forward-bias voltage. The impedance of the circuit in Figure \ref{fig:PN_eqcircuit}(b) is given by the following equation:
\begin{equation}
    Z = R_{s}+ \left[   R_{j}  \parallel \frac{1}{i \omega C_{j}}   \right] 
    \label{eq:PVimp}
\end{equation}
Equation \ref{eq:PVimp} can be rewritten into the form $Z=Z^{'}+iZ^{''}$ with the real part $Z^{'}$ and the imaginary part $Z^{''}$ being defined as follows:

\begin{equation}
    Z^{'} = R_{s} + \frac{R_{j}}{ 1 + R_{j}^{2} \omega^{2} C_{j}^{2}}
    \label{eq:PNimpreal}
\end{equation}

\begin{equation}
    Z^{''} = -\frac{R_{j}^{2} \omega C_{j}}{1 + R_{j}^{2} \omega^{2} C_{j}^{2}}
    \label{eq:PNimpcompl}
\end{equation}

In Equation \ref{eq:PNimpcompl}, the negative sign can be attributed to the behavior of current and voltage in capacitors. Specifically, in a capacitor, the current leads the voltage, resulting in a negative sign.

\section{Simulation approach}

In section \ref{sec:TCADmodel} the model used to generate impedance data is introduced. Subsequently, in section \ref{sec:data_analysis} it is explained how these impedance data are analyzed. 

\subsection{TCAD Sentaurus model} \label{sec:TCADmodel}
\begin{figure}[ht]
      \centering
      \includegraphics[width=0.41\textwidth]{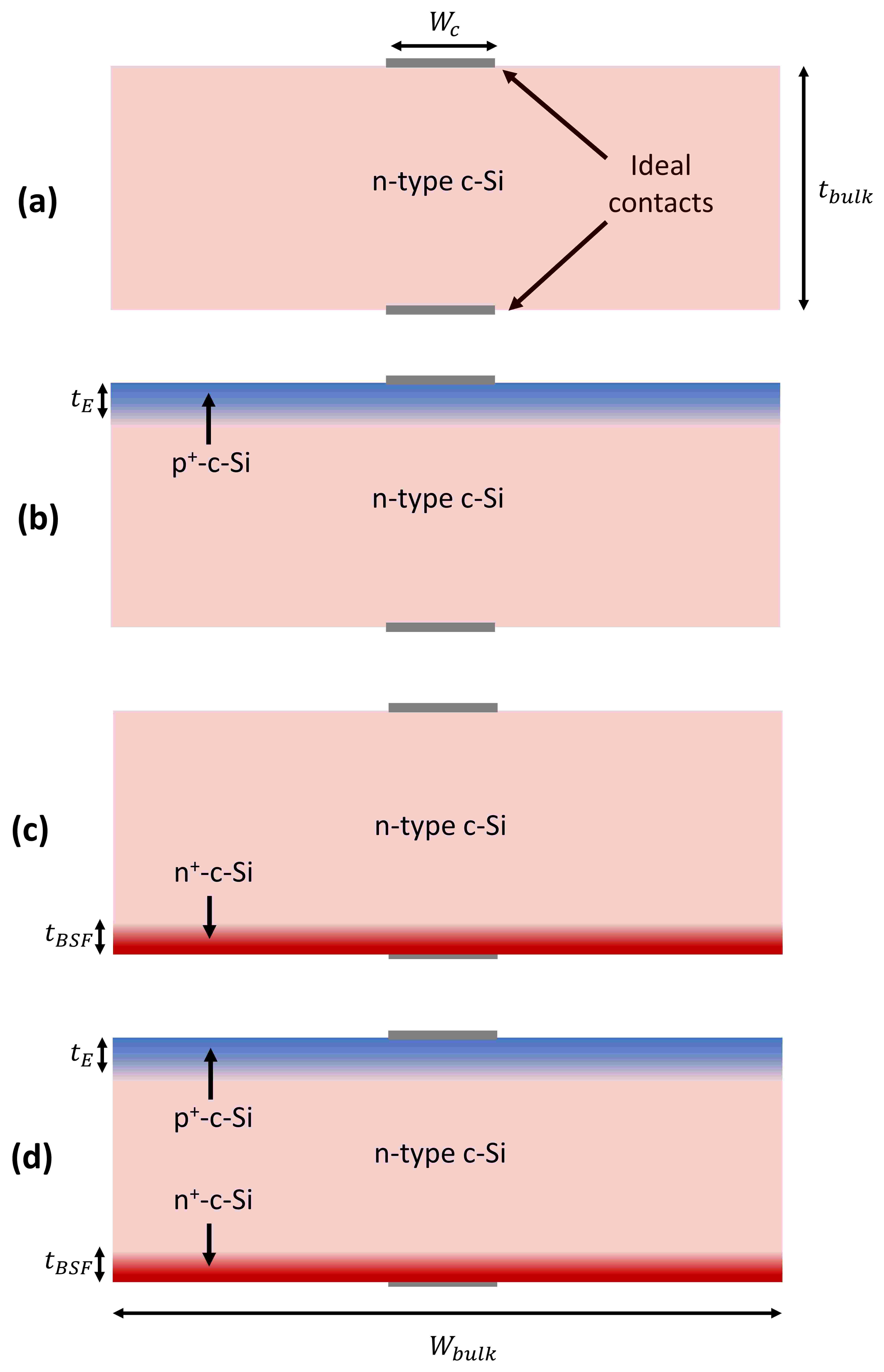}
      \caption{Simulated structures. Structure (a) consists of only an n-type bulk with two ideal contacts. Unless otherwise specified, in this study the bulk has a thickness $t_{\text{bulk}}$ of 200 $\mu$m, a width $W_{\text{bulk}}$ of 500 $\mu$m, and a bulk n-type dopant concentration $N_{\text{sub}}$ of 1 $\times$ 10\textsuperscript{15} cm\textsuperscript{-3}. The ideal contacts have a width $W_{\text{C}}$ of 50 $\mu$m. Structure (b) includes a p\textsuperscript{+} doped layer near the positive contact, called the emitter. Structure (c) includes an n\textsuperscript{+} doped layer near the negative contact, forming a so-called LH junction or BSF. Both the emitter and the BSF are composed of highly doped p\textsuperscript{+} and n\textsuperscript{+} regions, respectively, with Gaussian doping profiles. Each profile has a peak concentration of 1 $\times$ 10\textsuperscript{19} cm\textsuperscript{-3} at the surface, and drops to 10\% of the peak value at half of their respective junction depths $t_{\text{E}}$ and $t_{\text{BSF}}$, both set at 10 $\mu$m. Finally, structure (d) includes both a PN and LH junction.}
      \label{fig:PN_structure}
\end{figure}

In this study, electrical simulations of the semiconductor devices are performed using the finite element simulator TCAD Sentaurus \cite{SentaurusTCAD}. The two-dimensional device structures simulated in this work are illustrated in Figure \ref{fig:PN_structure}. The structure depicted in Figure \ref{fig:PN_structure}(a) consists of an n-type bulk and two ideal contacts with no resistance situated at the opposite side of the bulk. The top contact is from now on referred to as the positive contact, with the bottom contact being the negative contact. In this study different variations were investigated, but the results presented are primarily for a contact width of $W_{c}$ = 50 $\mu$m, a bulk width of $W_{\text{bulk}}$ = 500 $\mu$m, a thickness of $t_{\text{bulk}}$ = 200 $\mu$m, and a substrate/bulk n-type dopant concentration of $N_{\text{sub}}$ = 1 $\times$ 10\textsuperscript{15} cm\textsuperscript{-3}. 
The PN structure in Figure \ref{fig:PN_structure}(b) is obtained from the former by including a p\textsuperscript{+}-doped emitter directly underneath the positive contact. Alternatively, the LH structure in Figure \ref{fig:PN_structure}(c) includes an n\textsuperscript{+}-doped layer above the negative contact. The interface between this doped region and the bulk is also referred to as the low-high (LH) junction or the back surface field (BSF). It is worth noting that both the p\textsuperscript{+}-layer from Figure \ref{fig:PN_structure}(b) and the n\textsuperscript{+}- layer from Figure \ref{fig:PN_structure}(c) have Gaussian doping profiles with a peak concentration of 1 $\times$ 10\textsuperscript{19} cm\textsuperscript{-3} at the surface, which drops to 10\% of the peak value at half of their respective junction depths $t_{\text{E}}$ and $t_{\text{BSF}}$, both set at 10 $\mu$m. 
Finally, the PN/LH structure in Figure \ref{fig:PN_structure}(d) includes both an emitter and LH junction. 
Since it is expected from Equation \ref{eq:admittance_general} that the minority carrier lifetime plays an important role in the admittance, the trap density in the bulk is an important parameter that can impact both $C_{j}$ and $R_{j}$. In this study, the bulk trap density of the different structures in Figure \ref{fig:PN_structure} is adjusted using a parameter called the Shockley-Read-Hall (SRH) lifetime $\tau_{\text{SRH}}$.

The small-signal AC analysis is conducted using the \textit{ACCoupled} solve section of TCAD Sentaurus, assuming dark conditions and a temperature of 25 $^{\circ}$C. This simulation generates a frequency-dependent impedance matrix, comprising the real and imaginary parts of the impedance at each frequency examined during the AC analysis. In this study, the frequency range of interest spans from 10 $\mu$Hz to 1 GHz. 
It is worth noting that in experimental devices, the inductance of the metal contacts dominates the reactance at high frequencies \cite{vannijen2023}, whereas there is no such inductance in the simulation approach utilized here. Additionally, the use of ideal contacts in the simulation excludes any high-frequency effects that may alter the impedance of metal contacts. Moreover, the simulation data is characterized by low noise levels. Consequently, the simulation data enables the identification of subtle trends strictly related to the impedance of the junctions themselves, which may be more challenging to extract from experimental data.

\subsection{Equivalent circuit analysis} \label{sec:data_analysis}
The next step involves interpreting the impedance data using the electrical circuit depicted in Figure \ref{fig:PN_eqcircuit}(b). To extract the values of the different circuit elements, two different methods are employed:
\begin{enumerate}
    \item CNLS fitting: Complex nonlinear least-squares (CNLS) analysis is utilized to simultaneously fit Equations \ref{eq:PNimpreal} and \ref{eq:PNimpcompl} to the impedance data. This approach, widely applied in the analysis of c-Si solar cells, assumes fixed element values for $R_{j}$, $C_{j}$, and $R_{s}$. However, as explained in section \ref{sec:PNadmittance}, this assumption holds true only up to a certain frequency limit. This study further elaborates on the frequency limitations of this fixed circuit element assumption.
    \item Analytic approach: In the simulated device structures, $R_{s}$ is primarily influenced by the series resistance of the semiconductor bulk. It is generally assumed that $R_{s}$ is mostly independent of frequency \cite{Garland2011a}, and later in this work it is shown that this is also a valid approximation for the simulated device structures. Moreover, at high frequencies nearing 1 GHz, the real part of the impedance is primarily governed by $R_{s}$. Thus, one possible strategy is to extract the $R_{s}$ value directly from the high-frequency real part of the impedance. If the value of $R_{s}$ is known, $R_{j}$ and $C_{j}$ remain as the two unknowns in Equations \ref{eq:PNimpreal} and \ref{eq:PNimpcompl}. Their values can then be analytically computed from the impedance data at each individual frequency value. As such, this method facilitates the analysis of how $R_{j}$ and $C_{j}$ vary with frequency. It is worth noting that this analytic approach cannot be used for more complex equivalent circuits with more than two unknowns.

\end{enumerate}
As further elaborated in section \ref{sec:resultshighfreq}, throughout this study the low-frequency values of $R_{j}$ and $C_{j}$ from these two different methods closely align. However, differences between the methods emerge as the frequency exceeds a certain threshold.

\section{Results} \label{sec:resultshighfreq}
In section \ref{sec:PNhighfreq} the validity of the CNLS and analytic approach are compared over the whole tested frequency range. Subsequently, in section \ref{sec:bulkhighfreq} it is demonstrated how the silicon bulk properties influence these results. Finally, section \ref{sec:LHhighfreq} discusses how the impedance changes when a LH junction is added to the device structure. These three sections represent the main findings from this work. Design factors that have a more subtle influence on the impedance, such as variations in the dopant concentration and/or junction depth ($t_{\text{E}}$ and $t_{\text{BSF}}$) of the emitter and BSF, are not reported in this study.

\subsection{PN junction impedance at high frequency} \label{sec:PNhighfreq}
\begin{figure}[ht]
      \centering
      \includegraphics[width=0.41\textwidth]{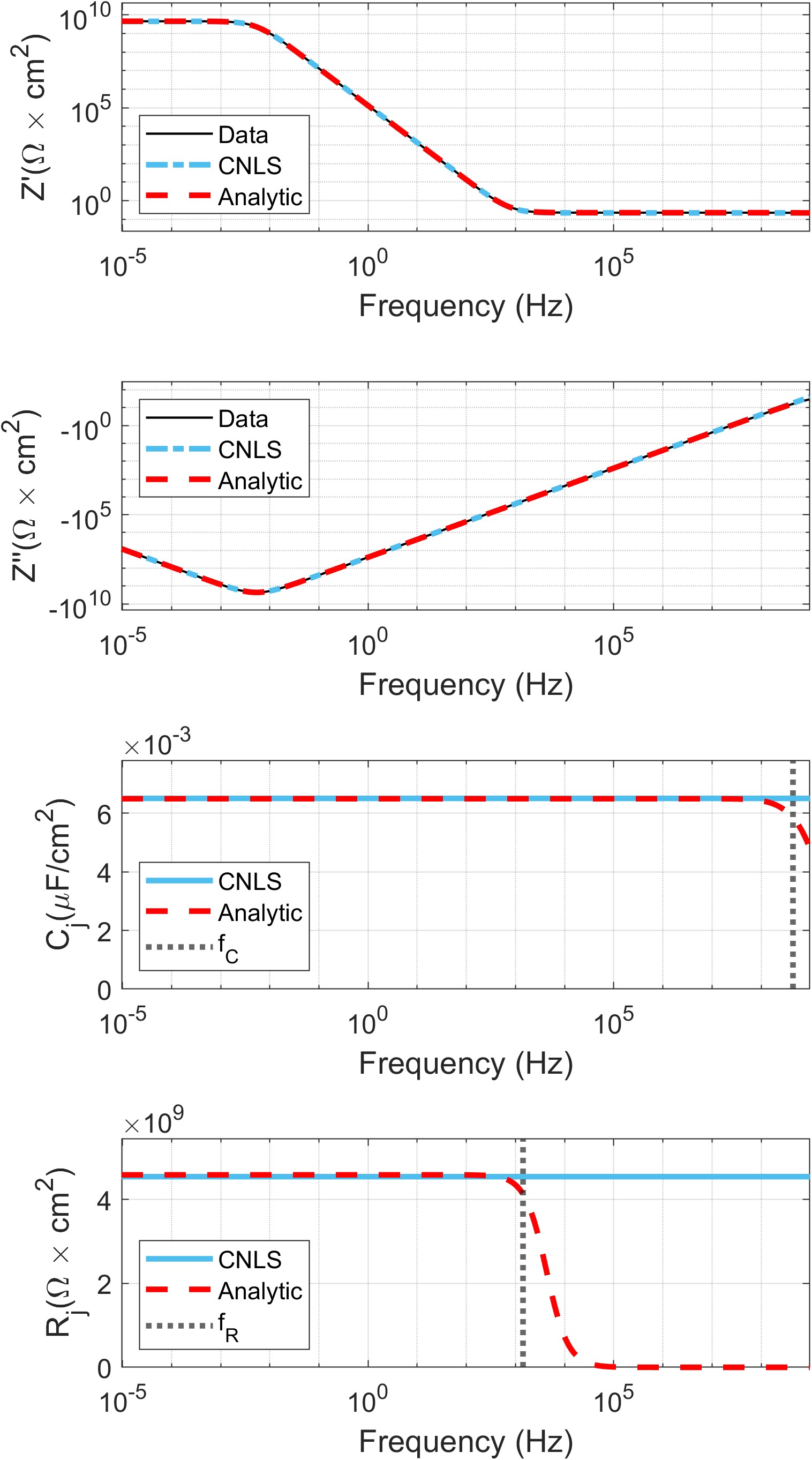}
      \caption{Impedance analysis of the PN structure of Figure \ref{fig:PN_structure}(b) at 0 mV bias voltage and $\tau_{\text{SRH}}$ = 10 ms. From top to bottom, the real part of the impedance, the imaginary part of the impedance, $C_{j}$ and $R_{j}$ are shown as functions of frequency. In the bottom two plots, $f_{C}$ = 444 MHz and $f_{R}$ = 1.4 kHz represent the frequencies where $C_{j}$ and $R_{j}$, respectively, drop below 90\% of their low-frequency value.}
      \label{fig:imp_000mV}
\end{figure}

\begin{figure}[ht]
      \centering
      \includegraphics[width=0.41\textwidth]{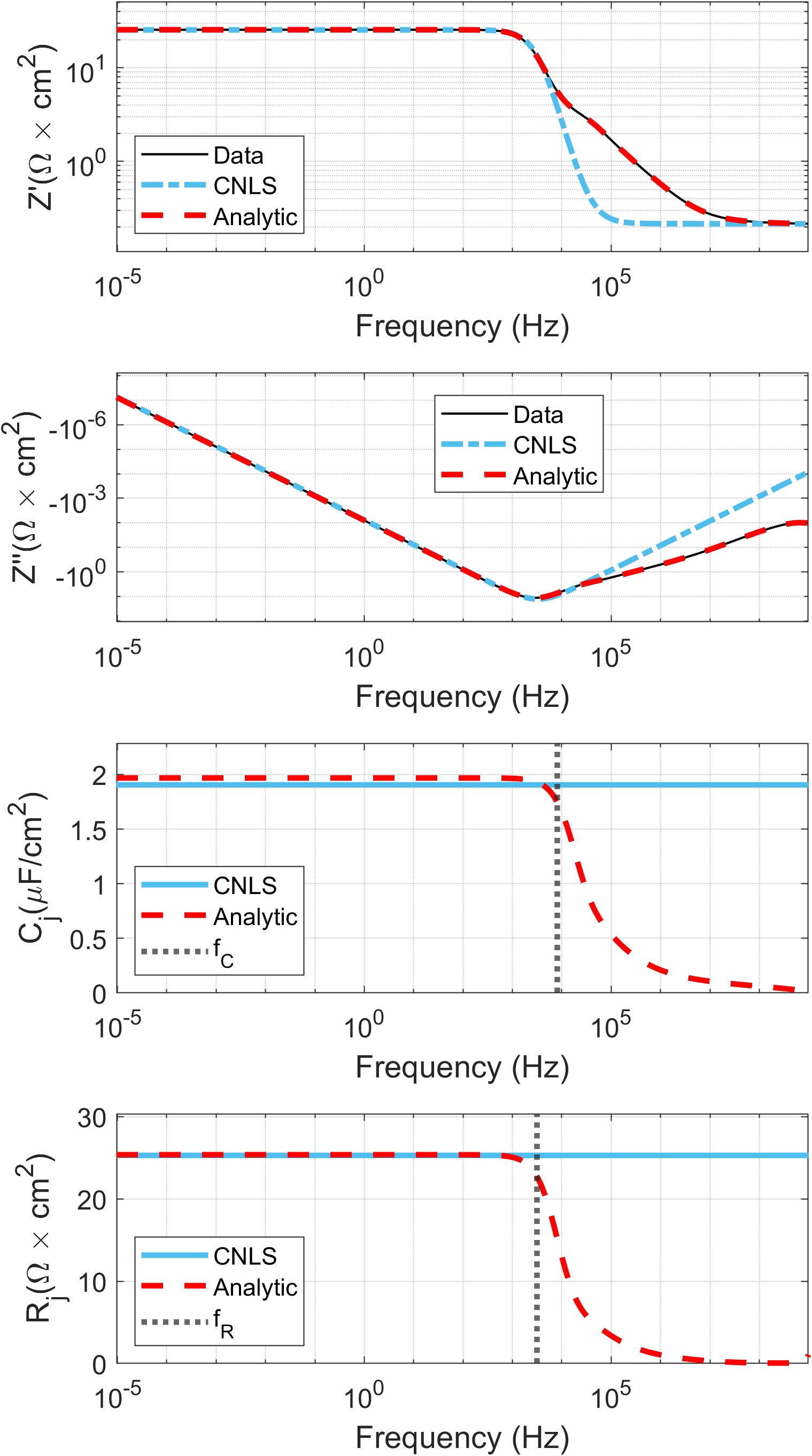}
      \caption{Impedance analysis of the PN structure of Figure \ref{fig:PN_structure}(b) at an applied forward bias voltage of 500 mV and $\tau_{\text{SRH}}$ = 10 ms. From top to bottom, the real part of the impedance, the imaginary part of the impedance, $C_{j}$ and $R_{j}$ are shown as functions of frequency. In the bottom two plots, $f_{C}$ = 8.1 kHz and $f_{R}$ = 3.1 kHz.}
      \label{fig:imp_500mV}
\end{figure}

In this section, the evolution of the PN junction impedance is discussed as the frequency reaches a level where the minority carrier concentration profile can no longer follow the AC signal. Previous literature suggests that the diffusion capacitance "contributes a frequency-dependent diffusion capacitance to the total capacitance" \cite{Green1973capacitance}. Additionally, it has been observed that the diffusion capacitance "relaxes" at high frequencies, meaning that its value reduces with frequency \cite{Lucia1993}. While discussions often focus on the frequency-dependent behavior of $C_{j}$, the frequency-dependence of $R_{j}$ is typically not discussed, potentially leading readers to assume that $R_{j}$ remains frequency-independent while $C_{j}$ decreases with frequency. In some works where a CPE is used in the equivalent model to address the capacitance variations, the resistive elements are typically fixed \cite{Shehata2023,Panigrahi2023}. However, it is worth pointing out that from Equations \ref{eq:admittance_general} and \ref{eq:admittance_math} it can be deduced that once the frequency exceeds a threshold where the condition $\omega \tau_{0}\ll1$ no longer holds, both the real and imaginary parts of the admittance become frequency-dependent. Specifically, both the real and imaginary parts of the admittance start to increase with frequency, indicating a relaxation of both $R_{j}$ and $C_{j}$. This theoretical prediction aligns with the TCAD simulation results from this study. Figures \ref{fig:imp_000mV} and \ref{fig:imp_500mV} depict the impedance data of the PN junction structure from Figure \ref{fig:PN_structure}(b) at 0 mV and 500 mV DC bias voltage, respectively. Moreover, these plots include the best CNLS and analytic fits. As expected, the fits from the analytic approach perfectly overlap with the data, since a point-by-point fitting is performed. Additionally, in the bottom two plots in Figures \ref{fig:imp_000mV} and \ref{fig:imp_500mV}, the $C_{j}$ and $R_{j}$ values extracted from both the CNLS and analytic methods are plotted as a function of frequency. 

Figure \ref{fig:imp_000mV} at 0 mV bias shows that the CNLS method offers a high-quality fit across the entire frequency range. Consequently, it might appear reasonable to conclude that $R_{j}$ and $C_{j}$ maintain fixed values throughout the frequency spectrum. Indeed, within the low-frequency range, the analytic approach yields identical values for $R_{j}$ and $C_{j}$ as the CNLS approach. However, at higher frequencies, the analytic approach indicates that $R_{j}$ and $C_{j}$ begin to relax within the tested frequency range. Here, $f_{C}$ and $f_{R}$ denote the frequencies where $C_{j}$ and $R_{j}$, respectively, have dropped below 90\% of their low-frequency value. In Figure \ref{fig:imp_000mV}, $f_{C}$ and $f_{R}$ are 444 MHz and 1.4 kHz, respectively. While the high value of $f_{C}$ can be attributed to the dominance of the depletion capacitance, the relatively low value of $f_{R}$ may seem unexpected, considering the high-quality CNLS fit. However, closer inspection reveals slight deviations between the CNLS fit and the impedance data, although these deviations are relatively minor and may not conclusively demonstrate that $R_{j}$ has indeed become variable.

Nevertheless, Figure \ref{fig:imp_500mV} at 500 mV forward bias provides deeper insights. Here, the deviations between the best CNLS fit and the impedance data clearly highlight the shortcomings of fixed circuit elements in accurately describing PN junction physics. The analytic approach suggests that $f_{R}$ remains similar to the 0 mV bias case, at 3.1 kHz. Moreover, at 500 mV bias, the anticipated variability of $C_{j}$ becomes noticeable from $f_{C}$ = 8.1 kHz and higher. The lower value of $f_{C}$ at 500 mV compared to 0 mV can be attributed to the dominance of the diffusion capacitance over $C_{j}$ in this case. In this study it was confirmed that no good fit can be achieved over the whole frequency range when $R_{j}$ is held fixed and $C_{j}$ is calculated at each individual frequency. Instead, the only plausible explanation is that both $R_{j}$ and $C_{j}$ relax with increasing frequency when $f_{R}$ and $f_{C}$ are exceeded, respectively. Finally, it is worth noting that in Figure \ref{fig:imp_500mV}, the low-frequency value of $C_{j}$ differs slightly between the CNLS and the analytic method. This difference arises because the CNLS fit is performed over the entire frequency range, also optimizing the fit beyond the frequency where $C_{j}$ and $R_{j}$ start to relax.

\subsection{Bulk properties influence on PN junction impedance} \label{sec:bulkhighfreq}
\begin{figure*}[ht]
      \centering
      \includegraphics[width=0.74\textwidth]{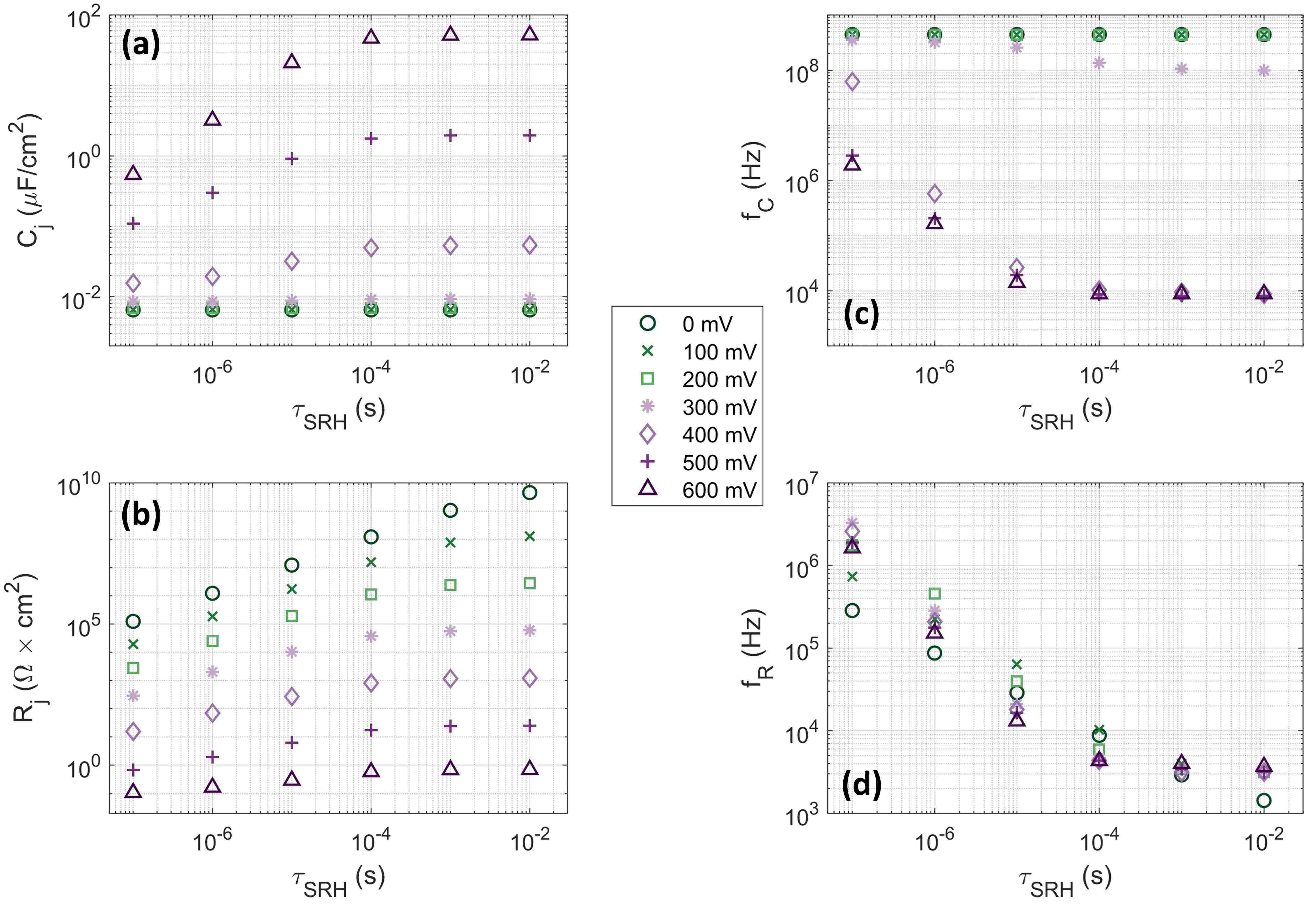}
      \caption{Equivalent circuit element analysis of the PN structure of Figure \ref{fig:PN_structure}(b). In (a) and (b), the low-frequency $C_{j}$ and $R_{j}$, which are determined using the analytic method, are presented as a function of $\tau_{\text{SRH}}$ for various DC bias voltages, respectively. Moreover, in (c) and (d), $f_{C}$ and $f_{R}$ are plotted as functions of $\tau_{\text{SRH}}$ for various DC bias voltages. $f_{C}$ and $f_{R}$ denote the frequencies where $C_{j}$ and $R_{j}$, respectively, have dropped below 90\% of their low-frequency value.}
      \label{fig:lifetime_wide}
\end{figure*}

In PN junctions where one side has a much higher dopant concentration than the other, the impedance is mainly affected by the properties of the lowly doped side of the junction. In the context of solar cells, it is thus important to mention that the substrate dopant concentration has a strong impact on the impedance \cite{vannijen2023}. However, since this effect is already well-known, the focus in this section shifts toward another bulk property, which is the bulk trap density.
Specifically, the influence of bulk trap density is investigated by varying $\tau_{\text{SRH}}$ in the PN junction structure depicted in Figure \ref{fig:PN_structure}(b). Figure \ref{fig:lifetime_wide}(a) illustrates how the low-frequency $C_{j}$ (determined using the analytic approach) varies as a function of $\tau_{\text{SRH}}$ for various applied DC bias voltages. 
At low bias voltage between 0 mV and 200 mV, where the depletion capacitance dominates, $C_{j}$ does not depend on $\tau_{\text{SRH}}$. This observation aligns with Equation \ref{eq:junccap}, which states that $C_{\text{dep}}$ is independent of the minority carrier lifetime.
However, at higher bias voltages, a different trend emerges. Specifically, in the range where $\tau_{\text{SRH}}$ is lower than 10\textsuperscript{-4} s, $C_{j}$ exhibits a strong increase with $\tau_{\text{SRH}}$. It is worth noting that the simulated $C_{j}$-$\tau_{\text{SRH}}$ relationship deviates from the square root dependency predicted by Equation \ref{eq:diffcap}. This deviation arises because, in the simulated device structure with a bulk thickness of 200 $\mu$m, a small fraction of carriers recombines at the ideal contacts rather than in the bulk. A reference simulation with a bulk thickness of 1000 $\mu$m and a $\tau_{\text{SRH}}$ variation between 10\textsuperscript{-7} s and 10\textsuperscript{-4} s (not shown in this paper) confirms this, as it exhibits the square root dependency predicted by Equation \ref{eq:diffcap}. 
Moreover, above $\tau_{\text{SRH}}$ = 10\textsuperscript{-4} s, the dependency of $C_{j}$ on $\tau_{\text{SRH}}$ diminishes. Presumably, this trend is attributed to the decreasing dominance of SRH recombination. As the hole diffusion length approaches or exceeds the bulk thickness, the effective lifetime in the bulk becomes dominated by minority carrier recombination at the ideal contacts. Indeed, as the effective hole lifetime increases from 10\textsuperscript{-5} s to 10\textsuperscript{-4} s, the theoretical relationship for the diffusion length ($L_{p} = \sqrt{D_{p} \tau_{p0}}$) predicts that $L_{p}$ increases from 110 $\mu$m to 349 $\mu$m, hereby exceeding the bulk thickness. 
In cases where the bulk thickness is much smaller than the minority carrier diffusion length, commonly observed in solar cells, the diode is referred to as a \textit{short diode}. The short diode explanation was further supported by confirming that for low $\tau_{\text{SRH}}$, $C_{j}$ does not depend on the bulk thickness, whereas for high $\tau_{\text{SRH}}$, $C_{j}$ does vary with the bulk thickness. However, these additional results are not included in this article.
Furthermore, Figure \ref{fig:lifetime_wide}(b) demonstrates the variation of $R_{j}$ with $\tau_{\text{SRH}}$. As expected, an increase in $\tau_{\text{SRH}}$ corresponds to higher $R_{j}$ values since the spatial distribution of minority carrier holes in the n-type bulk becomes less steep. Similar to the behavior observed for $C_{j}$ in Figure \ref{fig:lifetime_wide}(a), stabilization occurs as $\tau_{\text{SRH}}$ approaches levels where the short diode approximation applies. It is worth noting that this stabilization is not observed around 0 mV bias, presumably due to the dominance of generation and recombination currents at low voltages, unlike the diffusion current observed in forward bias \cite{Neamen2012}.

In Figure \ref{fig:lifetime_wide}(c), $f_{C}$ is plotted as functions of $\tau_{\text{SRH}}$. In the low-voltage region, where $C_{\text{dep}}$ dominates, $f_{C}$ remains relatively high and independent of $\tau_{\text{SRH}}$. However, in forward bias it can be observed that $f_{C}$ decreases with increasing $\tau_{\text{SRH}}$, with a stabilization occurring towards high $\tau_{\text{SRH}}$. Similarly, Figure \ref{fig:lifetime_wide}(d) demonstrates that $f_{R}$ varies with $\tau_{\text{SRH}}$, exhibiting a stabilization trend towards high $\tau_{\text{SRH}}$. This stabilization aligns with the earlier explanation that the impedance becomes less influenced by the bulk trap density as $\tau_{\text{SRH}}$ reaches a value where the short diode approximation becomes valid.

\begin{figure*}[ht]
      \centering
      \includegraphics[width=0.85\textwidth]{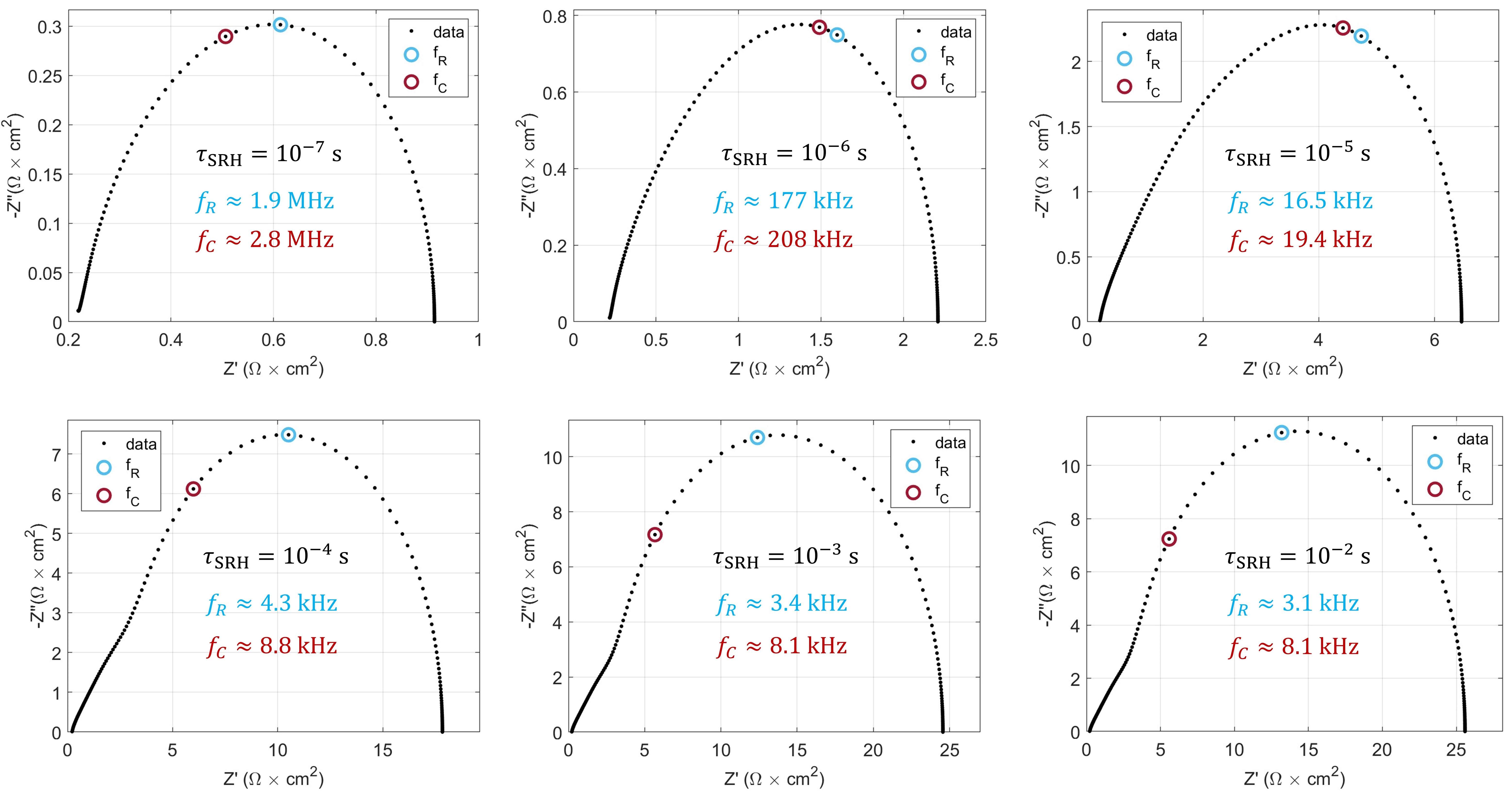}
      \caption{Nyquist spectra of the PN structure of Figure \ref{fig:PN_structure}(b) at 500 mV forward bias for varying $\tau_{\text{SRH}}$. $f_{C}$ and $f_{R}$ are highlighted in each plot.}
      \label{fig:Nyquist_PN_tau}
\end{figure*}

In Figure \ref{fig:Nyquist_PN_tau}, the Nyquist spectra of the PN junction devices at 500 mV forward bias are presented for varying $\tau_{\text{SRH}}$ values ranging from 10\textsuperscript{-7} s to 10\textsuperscript{-2} s. In each plot, $f_{R}$ and $f_{C}$ are highlighted.
Notably, $f_{R}$ and $f_{C}$ decrease as $\tau_{\text{SRH}}$ becomes higher. However, since the values of $R_{j}$ and $C_{j}$ change as well, $f_{R}$ and $f_{C}$ consistently fall within the frequency regime where the Nyquist spectra are near the top of the semi-circle. Consequently, all plots visibly deviate from the semi-circular shape expected of an $RC$-loop, indicating the frequency-dependence of $R_{j}$ and $C_{j}$. However, it must be noted that the shape of the Nyquist spectra may change when a LH junction is added at the negative contact. Further discussion on this topic is provided in the next section.

\subsection{Low-high junction influence} \label{sec:LHhighfreq}
\begin{figure}[ht]
      \centering
      \includegraphics[width=0.34\textwidth]{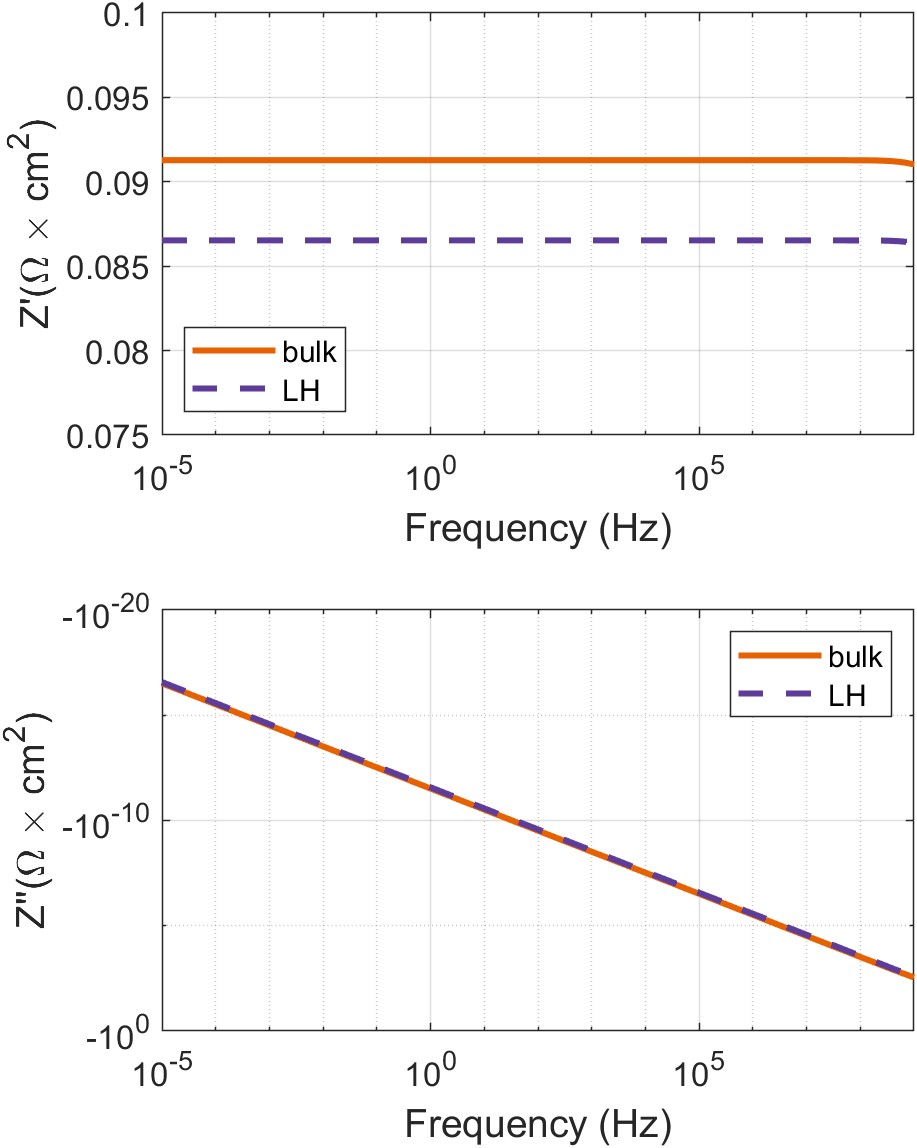}
      \caption{Impedance of the n-type bulk structure of Figure \ref{fig:PN_structure}(a) and the n-type bulk with n\textsuperscript{+}-doped LH structure of Figure \ref{fig:PN_structure}(c). $\tau_{\text{SRH}}$ = 10 ms and $W_{c}$ and $W_{\text{bulk}}$ are both equal to 500 $\mu$m. The plots are for 0 mV bias, but the results are nearly identical when a positive bias voltage is applied to the bulk contact with respect to the BSF contact.}
      \label{fig:bulkvsLH}
\end{figure}

In this section, the influence of a highly doped n\textsuperscript{+}-layer on the impedance is examined. As previously mentioned, practical c-Si solar cells commonly incorporate such a LH junction, also known as a back surface field (BSF). In addition to facilitating tunneling, the BSF serves as a barrier that prevents minority holes in the n-region from diffusing to the defective metal-semiconductor interface \cite{Smets2016}. Consequently, SRH recombination is reduced and the open-circuit voltage ($V_{oc}$) is enhanced. 
In literature on impedance spectroscopy for c-Si solar cells, the LH junction is often accounted for by incorporating an additional $RC$-loop into the equivalent circuit of the device \cite{Garland2011a,Crain2012,vannijen2023,Olayiwola2020,Panigrahi2016,Yadav2017}. However, there is limited discussion on comparing the impedance of PN junction devices with and without a BSF. Presumably, this is because practical devices inherently contain such a junction, making it challenging to experimentally test a device without it. 
With the simulation approach utilized in this study, further insights can be obtained about the impedance of the LH junction. Comparing an n-type bulk structure (Figure \ref{fig:PN_structure}(a)) with a similar structure incorporating a LH junction at the negative contact (Figure \ref{fig:PN_structure}(c)), the impedance spectra are plotted in Figure \ref{fig:bulkvsLH}.

In Figure \ref{fig:bulkvsLH} it is evident that the $Z'$ of the LH structure is slightly lower than that of the bulk structure, which is due to the higher conductivity in the n\textsuperscript{+} layer. Moreover, it can be observed from Figure \ref{fig:bulkvsLH} that both impedance spectra exhibit predominantly resistive behavior, with $Z'$ remaining constant and $Z''$ negligibly small across most frequencies. Hence, it is worth noting that this result confirms the validity of assuming a constant $R_{s}$ value in the analytic approach employed in this study. Only at the upper-frequency limit does $Z''$ become measurably large compared to $Z'$, leading to a -1.7$^{\circ}$ phase shift at 1 GHz. However, this high-frequency effect seems intrinsic to the bulk and not related to dynamic effects in the LH junction. In practical devices, inductance is likely to dominate $Z''$ at high frequencies \cite{vannijen2023}, obscuring such effects. Consequently, two key observations emerge. Firstly, the LH junction device structure exhibits lower $Z'$ compared to the bulk, meaning that the LH junction only lowers the series resistance and does not introduce additional resistive effects. Secondly, the LH junction itself does not seem to introduce dynamic effects within the frequency range of interest.

\begin{figure*}[ht]
      \centering
      \includegraphics[width=0.9\textwidth]{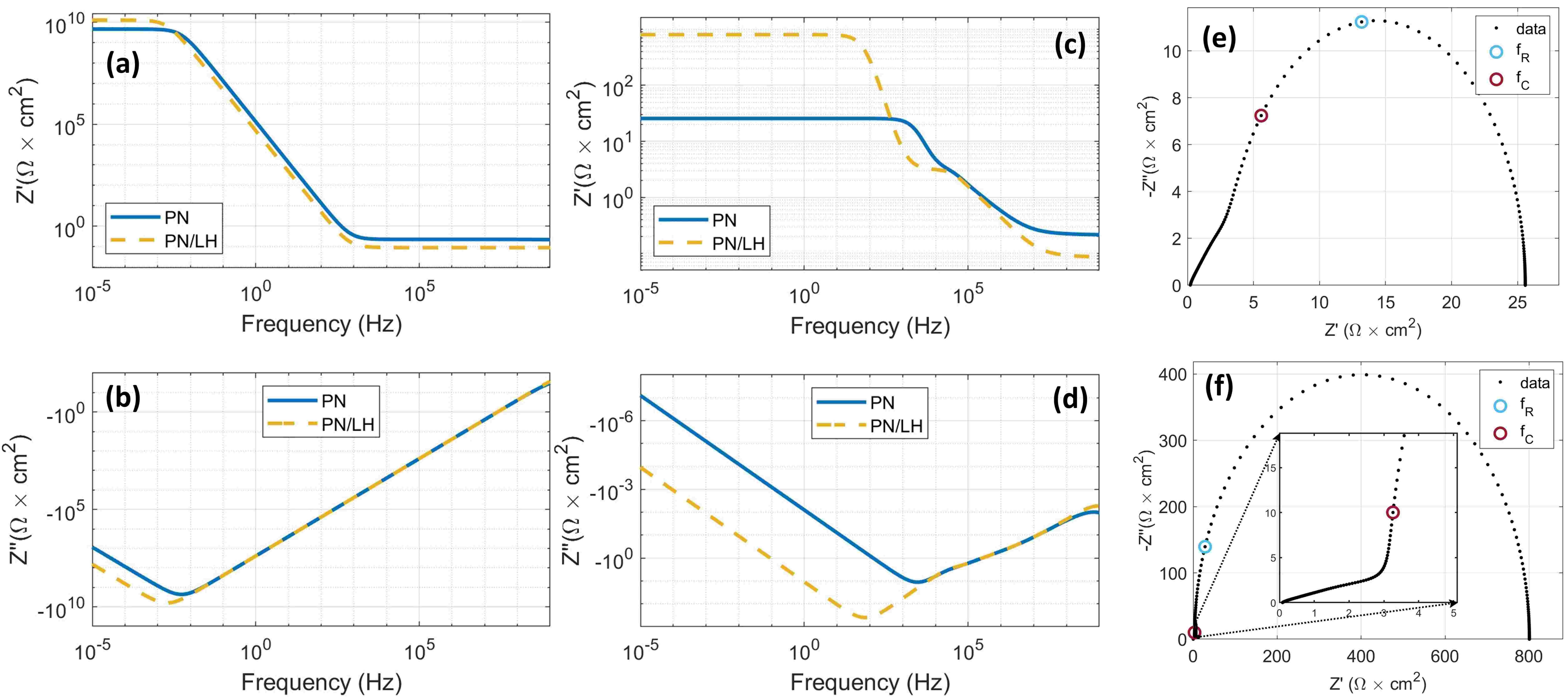}
      \caption{Comparison between the impedance of the PN structure of Figure \ref{fig:PN_structure}(b) and the PN/LH structure of Figure \ref{fig:PN_structure}(d) for $\tau_{\text{SRH}}$ = 10 ms. (a) and (b) present the impedance at 0 mV bias, whereas (c) and (d) are at 500 mV forward bias. Finally, (e) and (f) show the Nyquist spectra at 500 mV forward bias of the PN and PN/LH structure, respectively. For the PN structure at 500 mV bias, $f_{R}$ and $f_{C}$ are 3.1 kHz and 8.1 kHz, respectively. In contrast, for the PN/LH structure, $f_{R}$ and $f_{C}$ decrease to 0.40 kHz and 5.9 kHz, respectively.}
      \label{fig:PNvsPNLH}
\end{figure*}

In spite of the previous observations, the presence of a LH junction can significantly affect the impedance of a full PN junction device. To illustrate this, the impedance of (1) a PN device structure and (2) a PN/LH device structure are depicted in Figure \ref{fig:PNvsPNLH}. 
In the simulations it was observed that for both bias voltages, the electrostatic potential across the PN junction remains virtually unchanged between the device structures with and without the LH junction. This is related to the fact that the applied voltage mainly drops across the PN junction, rather than over the LH junction. Therefore, the differences in the impedance spectra due to addition of a LH junction are not related to changes in the bias voltage across the PN junction.

In the case of the 0 mV bias voltage depicted in Figure \ref{fig:PNvsPNLH}(a-b), there is a noticeable difference in the overall impedance between the two structures. Using the analytic approach, the differences in $R_{s}$ and the low-frequency values of $R_{j}$ and $C_{j}$ upon addition of the LH junction are compared. The low-frequency $R_{j}$ increases from 4.54 $\times 10^{9}$ $\Omega \text{cm}^{2}$ to 1.25 $\times 10^{10}$ $\Omega \text{cm}^{2}$, while $R_{s}$ decreases from 0.22 $\Omega \text{cm}^{2}$ to 0.088 $\Omega \text{cm}^{2}$. The increase in $R_{j}$ can primarily be attributed to the fact that the BSF repels minority carriers from the negative contact. Consequently, the distribution of minority carriers throughout the bulk becomes less steep, resulting in an increase in $R_{j}$. Conversely, the decreased $R_{s}$ can be attributed to the enhanced lateral charge carrier transport within the highly doped region of the LH junction. It is worth noting that the low-frequency $C_{j}$ remains unchanged at 6.50 nF/cm\textsuperscript{2} between the two device structures. This consistency can be attributed to the fact that the depletion capacitance solely relies on the ionized atoms within the depletion region of the PN junction.

For the 500 mV case in Figure \ref{fig:PNvsPNLH}(c-d), the addition of the LH junction increases the low-frequency $R_{j}$ from 25.3 $\Omega \text{cm}^{2}$ to 801 $\Omega \text{cm}^{2}$. $R_{s}$ is again decreased from 0.22 $\Omega \text{cm}^{2}$ to 0.088 $\Omega \text{cm}^{2}$. The explanations are similar to the 0 mV case. However, opposed to the 0 mV case, the addition of the LH junction at 500 mV increases the low-frequency $C_{j}$ from 1.97 $\mu$F/cm\textsuperscript{2} to 2.72 $\mu$F/cm\textsuperscript{2}. This increase can be attributed to the fact that the BSF repels the minority carrier holes from the negative contact, increasing the effective lifetime of holes stored in the silicon bulk. Furthermore, it is worth reflecting how the addition of the LH junction affects the Nyquist plots. The Nyquist spectra at 500 mV bias are presented in Figure \ref{fig:PNvsPNLH}(e) for the PN device, and in Figure \ref{fig:PNvsPNLH}(f) for the PN/LH device. At first glance, the Nyquist spectrum of the device with the PN/LH junction seems to closely resemble a semi-circular shape, as opposed to the PN structure. However, the zoomed-in view in Figure \ref{fig:PNvsPNLH}(f) shows that the high-frequency variations of $R_{j}$ and $C_{j}$ still occur and create distortions to the high-frequency tail of the Nyquist spectrum.

\begin{figure}[ht]
      \centering
      \includegraphics[width=0.4\textwidth]{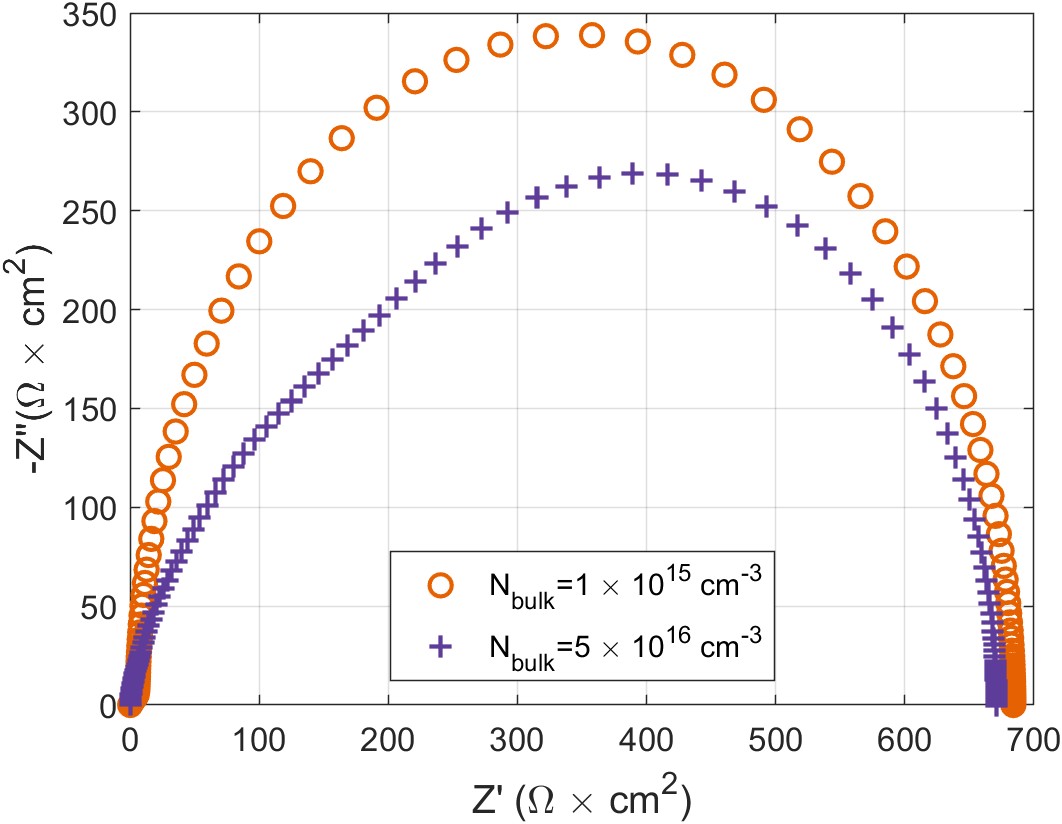}
      \caption{Nyquist spectra of the PN/LH structure of Figure \ref{fig:PN_structure}(d) for wafer dopant concentration values of $N_{\text{sub}}$=1 $\times$ 10\textsuperscript{15} cm\textsuperscript{-3} and $N_{\text{sub}}$=5 $\times$ 10\textsuperscript{16} cm\textsuperscript{-3}. 
      The impedance data are recorded at 500 mV forward bias with $\tau_{\text{SRH}}$ = 10 ms.}
      \label{fig:PNLH_Nyquist_compare}
\end{figure}

In this section, it was demonstrated that the LH junction exhibits no significant impedance characteristics by itself. This finding contrasts with existing literature, where the LH junction is often represented as an additional $RC$-loop in series with the $R_{j}$-$C_{j}$ loop \cite{Garland2011a,Crain2012,vannijen2023,Olayiwola2020,Panigrahi2016,Yadav2017}. Indeed, particular combinations of parameters in devices with a PN/LH structure can result in a Nyquist plot resembling that of two $RC$-loops in series. For instance, Figure \ref{fig:PNLH_Nyquist_compare} illustrates how the Nyquist spectra of a PN/LH device changes when the bulk dopant concentration is changed between $N_{\text{sub}}$=1 $\times$ 10\textsuperscript{15} and $N_{\text{sub}}$=5 $\times$ 10\textsuperscript{16} cm\textsuperscript{-3}. While one resembles a single semicircle, the other might be mistaken for two superimposed semicircles. However, this study suggests that the inclusion of an $R_{LH}$-$C_{LH}$ loop to the model does not accurately represent the underlying physics. Instead, its inclusion in the model appears to account for the frequency-dependent variations in $R_{j}$ and $C_{j}$. Throughout our investigation, instances were observed where the two $RC$-loop model falls short in explaining the impedance spectra adequately, indicating that the relaxation of $R_{j}$ and $C_{j}$ is a more plausible explanation. Moreover, such relaxation phenomena cannot always be easily detected through Nyquist plots on linear scales. Therefore, Bode plots or real and imaginary impedance plots on a logarithmic scale often provide better insights.

\section{Conclusion}
The impedance characteristics of PN homojunction devices were investigated, particularly focusing on the effects of frequency, applied bias voltage, bulk trap density, and the presence of a low-high (LH) junction or back surface field (BSF). The employed TCAD simulation method allows to detect subtle trends that may be difficult to identify experimentally due to noise and reactance of metal contacts. Through analysis of the impedance data it was revealed that the PN junction exhibits a fixed $RC$-loop behavior at low frequencies, but undergoes relaxation in both resistance $R_{j}$ and capacitance $C_{j}$ as frequency increases. 
Additionally, it was found that when the bulk defect density decreases below a certain threshold, the effective minority carrier lifetime in the bulk becomes limited by recombination at the contacts. In this regime, the short diode approximation effectively explains various observed impedance trends and factors such as wafer thickness and the presence of a BSF strongly affect the impedance. Notably, it was shown that the addition of a LH junction repels minority carriers from this contact, hereby impacting the impedance by altering $R_{j}$, $C_{j}$, and $R_{s}$. Contrary to conventional modeling approaches, which often include an additional $RC$-loop to represent the LH junction, this study suggests that such a representation does not represent the underlying physics, particularly the frequency-dependent behavior of $R_{j}$ and $C_{j}$. 
These insights are important for solar cell applications that require a thorough understanding of the small-signal response. Moreover, this work serves as a basis for fully understanding the impedance of solar cells with passivating contacts.




\printcredits

\subsection*{Acknowledgements}
This work is supported by the sector plan of the Dutch government in photovoltatronics research. 

\subsection*{Declaration of generative AI and AI-assisted technologies in the writing process}
During the preparation of this work the author(s) used ChatGPT in order to improve the text flow. After using this tool/service, the author(s) reviewed and edited the content as needed and take(s) full responsibility for the content of the published article.

\subsection*{Data availability}
The data for this article will be made available at the 4TU Research Database, accessible with this link: \url{https://doi.org/10.4121/19445ed8-c8d5-4204-a8c6-adaa7a55ece3}

\bibliographystyle{model1-num-names}

\bibliography{mybibfile}

\begin{thebibliography}{22}
\expandafter\ifx\csname natexlab\endcsname\relax\def\natexlab#1{#1}\fi
\providecommand{\url}[1]{\texttt{#1}}
\providecommand{\href}[2]{#2}
\providecommand{\path}[1]{#1}
\providecommand{\DOIprefix}{doi:}
\providecommand{\ArXivprefix}{arXiv:}
\providecommand{\URLprefix}{URL: }
\providecommand{\Pubmedprefix}{pmid:}
\providecommand{\doi}[1]{\href{http://dx.doi.org/#1}{\path{#1}}}
\providecommand{\Pubmed}[1]{\href{pmid:#1}{\path{#1}}}
\providecommand{\bibinfo}[2]{#2}
\ifx\xfnm\relax \def\xfnm[#1]{\unskip,\space#1}\fi
\bibitem[{Mora-Seró et~al.(2009)Mora-Seró, Garcia-Belmonte, Boix, Vázquez, and Bisquert}]{Mora2009}
\bibinfo{author}{I.~Mora-Seró}, \bibinfo{author}{G.~Garcia-Belmonte}, \bibinfo{author}{P.~P. Boix}, \bibinfo{author}{M.~A. Vázquez}, \bibinfo{author}{J.~Bisquert},
\newblock \bibinfo{title}{Impedance spectroscopy characterisation of highly efficient silicon solar cells under different light illumination intensities},
\newblock \bibinfo{journal}{Energy Environ. Sci.} \bibinfo{volume}{2} (\bibinfo{year}{2009}) \bibinfo{pages}{678--686}.
\bibitem[{Garland et~al.(2011)Garland, Crain, Zheng, Sulyma, and Roy}]{Garland2011a}
\bibinfo{author}{J.~E. Garland}, \bibinfo{author}{D.~J. Crain}, \bibinfo{author}{J.~P. Zheng}, \bibinfo{author}{C.~M. Sulyma}, \bibinfo{author}{D.~Roy},
\newblock \bibinfo{title}{Electro-analytical characterization of photovoltaic cells by combining voltammetry and impedance spectroscopy: voltage dependent parameters of a silicon solar cell under controlled illumination and temperature},
\newblock \bibinfo{journal}{Energy Environ. Sci.} \bibinfo{volume}{4} (\bibinfo{year}{2011}) \bibinfo{pages}{485--498}.
\bibitem[{Mora-Seró et~al.(2008)Mora-Seró, Luo, Garcia-Belmonte, Bisquert, Muñoz, Voz, Puigdollers, and Alcubilla}]{MoraSero2008}
\bibinfo{author}{I.~Mora-Seró}, \bibinfo{author}{Y.~Luo}, \bibinfo{author}{G.~Garcia-Belmonte}, \bibinfo{author}{J.~Bisquert}, \bibinfo{author}{D.~Muñoz}, \bibinfo{author}{C.~Voz}, \bibinfo{author}{J.~Puigdollers}, \bibinfo{author}{R.~Alcubilla},
\newblock \bibinfo{title}{Recombination rates in heterojunction silicon solar cells analyzed by impedance spectroscopy at forward bias and under illumination},
\newblock \bibinfo{journal}{Solar Energy Materials and Solar Cells} \bibinfo{volume}{92} (\bibinfo{year}{2008}) \bibinfo{pages}{505--509}.
\bibitem[{Panigrahi et~al.(2016)Panigrahi, Vandana, Singh, Batra, Gope, Sharma, Pathi, Srivastava, Rauthan, and Singh}]{Panigrahi2016}
\bibinfo{author}{J.~Panigrahi}, \bibinfo{author}{Vandana}, \bibinfo{author}{R.~Singh}, \bibinfo{author}{N.~Batra}, \bibinfo{author}{J.~Gope}, \bibinfo{author}{M.~Sharma}, \bibinfo{author}{P.~Pathi}, \bibinfo{author}{S.~Srivastava}, \bibinfo{author}{C.~Rauthan}, \bibinfo{author}{P.~Singh},
\newblock \bibinfo{title}{Impedance spectroscopy of crystalline silicon solar cell: Observation of negative capacitance},
\newblock \bibinfo{journal}{Solar Energy} \bibinfo{volume}{136} (\bibinfo{year}{2016}) \bibinfo{pages}{412--420}.
\bibitem[{{van Nijen} et~al.(2022){van Nijen}, Manganiello, Zeman, and Isabella}]{vannijen2022}
\bibinfo{author}{D.~A. {van Nijen}}, \bibinfo{author}{P.~Manganiello}, \bibinfo{author}{M.~Zeman}, \bibinfo{author}{O.~Isabella},
\newblock \bibinfo{title}{{Exploring the benefits, challenges, and feasibility of integrating power electronics into c-Si solar cells}},
\newblock \bibinfo{journal}{Cell Reports Physical Science} \bibinfo{volume}{3} (\bibinfo{year}{2022}) \bibinfo{pages}{100944}.
\bibitem[{Huang et~al.(2016)Huang, Lehman, and Qian}]{Huang2016}
\bibinfo{author}{J.-H. Huang}, \bibinfo{author}{B.~Lehman}, \bibinfo{author}{T.~Qian},
\newblock \bibinfo{title}{{Submodule integrated boost DC-DC converters with no external input capacitor or input inductor for low power photovoltaic applications}},
\newblock in: \bibinfo{booktitle}{2016 IEEE Energy Conversion Congress and Exposition (ECCE)}, \bibinfo{year}{2016}, pp. \bibinfo{pages}{1--7}. \DOIprefix\doi{10.1109/ECCE.2016.7855476}.
\bibitem[{Ziar et~al.(2021)Ziar, Manganiello, Isabella, and Zeman}]{Ziar2021}
\bibinfo{author}{H.~Ziar}, \bibinfo{author}{P.~Manganiello}, \bibinfo{author}{O.~Isabella}, \bibinfo{author}{M.~Zeman},
\newblock \bibinfo{title}{{Photovoltatronics: intelligent PV-based devices for energy and information applications}},
\newblock \bibinfo{journal}{Energy Environ. Sci.} \bibinfo{volume}{14} (\bibinfo{year}{2021}) \bibinfo{pages}{106--126}.
\bibitem[{Kim et~al.(2014)Kim, Won, and Nahm}]{kim2014simultaneous}
\bibinfo{author}{S.-M. Kim}, \bibinfo{author}{J.-S. Won}, \bibinfo{author}{S.-H. Nahm},
\newblock \bibinfo{title}{Simultaneous reception of solar power and visible light communication using a solar cell},
\newblock \bibinfo{journal}{Optical Engineering} \bibinfo{volume}{53} (\bibinfo{year}{2014}) \bibinfo{pages}{046103--046103}.
\bibitem[{Sarwar et~al.(2017)Sarwar, Sun, Kong, Ali, Yu, Cong, and Xu}]{RSarwar2017}
\bibinfo{author}{R.~Sarwar}, \bibinfo{author}{B.~Sun}, \bibinfo{author}{M.~Kong}, \bibinfo{author}{T.~Ali}, \bibinfo{author}{C.~Yu}, \bibinfo{author}{B.~Cong}, \bibinfo{author}{J.~Xu},
\newblock \bibinfo{title}{Visible light communication using a solar-panel receiver},
\newblock in: \bibinfo{booktitle}{2017 16th International Conference on Optical Communications and Networks (ICOCN)}, \bibinfo{year}{2017}, pp. \bibinfo{pages}{1--3}. \DOIprefix\doi{10.1109/ICOCN.2017.8121577}.
\bibitem[{Zhou et~al.(2023)Zhou, Ibrahim, Muttillo, Manganiello, Ziar, and Isabella}]{YZhou2023}
\bibinfo{author}{Y.~Zhou}, \bibinfo{author}{A.~Ibrahim}, \bibinfo{author}{M.~Muttillo}, \bibinfo{author}{P.~Manganiello}, \bibinfo{author}{H.~Ziar}, \bibinfo{author}{O.~Isabella},
\newblock \bibinfo{title}{{Bandwidth Characterization of c-Si Solar Cells as VLC Receiver under Colored LEDs}},
\newblock in: \bibinfo{booktitle}{{2023 8th International Conference on Smart and Sustainable Technologies (SpliTech)}}, \bibinfo{year}{2023}, pp. \bibinfo{pages}{1--5}. \DOIprefix\doi{10.23919/SpliTech58164.2023.10193624}.
\bibitem[{Neamen(2012)}]{Neamen2012}
\bibinfo{author}{D.~A. Neamen}, \bibinfo{title}{{Semiconductor Physics and Devices}}, \bibinfo{edition}{4} ed., \bibinfo{publisher}{McGraw Hill Education}, \bibinfo{year}{2012}.
\bibitem[{Allen et~al.(2019)Allen, Bullock, Yang, Javey, and Wolf}]{Allen2019}
\bibinfo{author}{T.~G. Allen}, \bibinfo{author}{J.~Bullock}, \bibinfo{author}{X.~Yang}, \bibinfo{author}{A.~Javey}, \bibinfo{author}{S.~D. Wolf},
\newblock \bibinfo{title}{{Passivating contacts for crystalline silicon solar cells}},
\newblock \bibinfo{journal}{Nature Energy} \bibinfo{volume}{4} (\bibinfo{year}{2019}) \bibinfo{pages}{914--928}.
\bibitem[{Crain et~al.(2012)Crain, Garland, Rock, and Roy}]{Crain2012}
\bibinfo{author}{D.~J. Crain}, \bibinfo{author}{J.~E. Garland}, \bibinfo{author}{S.~E. Rock}, \bibinfo{author}{D.~Roy},
\newblock \bibinfo{title}{Quantitative characterization of silicon solar cells in the electro-analytical approach: Combined measurements of temperature and voltage dependent electrical parameters},
\newblock \bibinfo{journal}{Anal. Methods} \bibinfo{volume}{4} (\bibinfo{year}{2012}) \bibinfo{pages}{106--117}.
\bibitem[{{van Nijen} et~al.(2023){van Nijen}, Muttillo, {Van Dyck}, Poortmans, Zeman, Isabella, and Manganiello}]{vannijen2023}
\bibinfo{author}{D.~A. {van Nijen}}, \bibinfo{author}{M.~Muttillo}, \bibinfo{author}{R.~{Van Dyck}}, \bibinfo{author}{J.~Poortmans}, \bibinfo{author}{M.~Zeman}, \bibinfo{author}{O.~Isabella}, \bibinfo{author}{P.~Manganiello},
\newblock \bibinfo{title}{{Revealing capacitive and inductive effects in modern industrial c-Si photovoltaic cells through impedance spectroscopy}},
\newblock \bibinfo{journal}{Solar Energy Materials and Solar Cells} \bibinfo{volume}{260} (\bibinfo{year}{2023}) \bibinfo{pages}{112486}.
\bibitem[{Olayiwola and Barendse(2020)}]{Olayiwola2020}
\bibinfo{author}{O.~I. Olayiwola}, \bibinfo{author}{P.~S. Barendse},
\newblock \bibinfo{title}{{Photovoltaic Cell/Module Equivalent Electric Circuit Modeling Using Impedance Spectroscopy}},
\newblock \bibinfo{journal}{IEEE Transactions on Industry Applications} \bibinfo{volume}{56} (\bibinfo{year}{2020}) \bibinfo{pages}{1690--1701}.
\bibitem[{Yadav et~al.(2017)Yadav, Pandey, Bhatt, Kumar, and Kim}]{Yadav2017}
\bibinfo{author}{P.~Yadav}, \bibinfo{author}{K.~Pandey}, \bibinfo{author}{V.~Bhatt}, \bibinfo{author}{M.~Kumar}, \bibinfo{author}{J.~Kim},
\newblock \bibinfo{title}{{Critical aspects of impedance spectroscopy in silicon solar cell characterization: A review}},
\newblock \bibinfo{journal}{Renewable and Sustainable Energy Reviews} \bibinfo{volume}{76} (\bibinfo{year}{2017}) \bibinfo{pages}{1562--1578}.
\bibitem[{Shehata et~al.(2023)Shehata, Truong, Basnet, Nguyen, Macdonald, and Black}]{Shehata2023}
\bibinfo{author}{M.~M. Shehata}, \bibinfo{author}{T.~N. Truong}, \bibinfo{author}{R.~Basnet}, \bibinfo{author}{H.~T. Nguyen}, \bibinfo{author}{D.~H. Macdonald}, \bibinfo{author}{L.~E. Black},
\newblock \bibinfo{title}{{Impedance spectroscopy characterization of c-Si solar cells with SiO\textsubscript{x}/ Poly-Si rear passivating contacts}},
\newblock \bibinfo{journal}{Solar Energy Materials and Solar Cells} \bibinfo{volume}{251} (\bibinfo{year}{2023}) \bibinfo{pages}{112167}.
\bibitem[{Panigrahi et~al.(2023)Panigrahi, Pandey, Bhattacharya, Pal, Mandal, and {Krishna Komarala}}]{Panigrahi2023}
\bibinfo{author}{J.~Panigrahi}, \bibinfo{author}{A.~Pandey}, \bibinfo{author}{S.~Bhattacharya}, \bibinfo{author}{A.~Pal}, \bibinfo{author}{S.~Mandal}, \bibinfo{author}{V.~{Krishna Komarala}},
\newblock \bibinfo{title}{{Impedance spectroscopy of amorphous/crystalline silicon heterojunction solar cells under dark and illumination}},
\newblock \bibinfo{journal}{Solar Energy} \bibinfo{volume}{259} (\bibinfo{year}{2023}) \bibinfo{pages}{165--173}.
\bibitem[{Lucia et~al.(1993)Lucia, Hernandez-Rojas, Leon, and Mártil}]{Lucia1993}
\bibinfo{author}{M.~L. Lucia}, \bibinfo{author}{J.~L. Hernandez-Rojas}, \bibinfo{author}{C.~Leon}, \bibinfo{author}{I.~Mártil},
\newblock \bibinfo{title}{{Capacitance measurements of p-n junctions: depletion layer and diffusion capacitance contributions}},
\newblock \bibinfo{journal}{European Journal of Physics} \bibinfo{volume}{14} (\bibinfo{year}{1993}) \bibinfo{pages}{86}.
\bibitem[{Synopsis(2021)}]{SentaurusTCAD}
\bibinfo{author}{Synopsis}, \bibinfo{title}{Sentaurus device user}, \bibinfo{year}{2021}.
\bibitem[{Green and Gunn(1973)}]{Green1973capacitance}
\bibinfo{author}{M.~Green}, \bibinfo{author}{M.~Gunn},
\newblock \bibinfo{title}{{The capacitance of abrupt p-n junction diodes under forward bias}},
\newblock \bibinfo{journal}{physica status solidi (a)} \bibinfo{volume}{19} (\bibinfo{year}{1973}) \bibinfo{pages}{K93--K96}.
\bibitem[{Smets et~al.(2016)Smets, J{\"{a}}ger, Isabella, van Swaaij, and Zeman}]{Smets2016}
\bibinfo{author}{A.~Smets}, \bibinfo{author}{K.~J{\"{a}}ger}, \bibinfo{author}{O.~Isabella}, \bibinfo{author}{R.~van Swaaij}, \bibinfo{author}{M.~Zeman}, \bibinfo{title}{{Solar Energy - The Physics and Engineering of Photovoltaic Conversion Technologies and Systems}}, \bibinfo{publisher}{UIT Cambridge Ltd}, \bibinfo{year}{2016}.

\end{thebibliography}



\end{document}